\documentclass[%
 reprint,
 amsmath,amssymb,
 aps,
]{revtex4-1}

\usepackage{graphicx}% Include figure files
\usepackage{dcolumn}% Align table columns on decimal point
\usepackage{bm}% bold math
\begin{document}

\title{Gravitational Wave Formation from the Collapse of Dark Energy Field Configurations}% Force line breaks with \\

\author{Vishal Jhalani, Ayushi Mishra and Anupam Singh}
 
 \affiliation{Physics Department, L.N. Mittal I.I.T, Jaipur, Rajasthan, India.}

\date{\today}% It is always \today, today,

\begin{abstract}
Dark Energy is the dominant component of the energy density of the universe. 
In a previous paper, we have shown that the collapse of dark energy fields leads to the formation of Super Massive Black Holes with masses comparable to the masses of Black Holes at the centers of galaxies.
Thus it becomes a pressing issue to investigate the other physical consequences of the collapse of Dark Energy fields. Given that the primary interactions of Dark Energy fields with the rest of the Universe are gravitational, it is particularly interesting to investigate the gravitational wave signals emitted during the process of the collapse of Dark Energy fields. This is the focus of the current work described in this paper.
We describe and use the 3+1 BSSN formalism to follow the evolution of the dark energy fields coupled with gravity and to extract the gravitational wave signals. Finally, we describe the results of our numerical computations and the gravitational wave signals produced as a result of the collapse of the dark energy fields. 

\end{abstract}

\pacs{Valid PACS appear here}% PACS, the Physics and Astronomy

\maketitle

\section{INTRODUCTION}

%\tableofcontents

As the accuracy of astronomical observations has increased which in turn has led us to an
increased precision in the determination of cosmological parameters.
This, in turn, led us to critically re-examine our cosmological models.
In particular,  the accurate determination of the Hubble constant and
the independent determination of the age of the universe has forced us to
critically re-examine the simplest cosmological model
- a flat universe with a zero cosmological constant\cite{Pierce,Freedman}.
These observations forced one of the current authors\cite{Singh} to consider the idea of
a small non-vanishing vacuum energy due to fields as playing an important role in the Universe.
Subsequently, there has been a large body of work both on the observational and theoretical side
that has firmed up our belief in what we now call dark energy.

Given that we now believe that dark energy is the dominant component of the Universe, it is now a pressing need
to understand the dynamics of dark energy and in particular the gravitational dynamics of dark energy.

Before we start our detailed look at the gravitational dynamics of dark energy field configurations we would like to
quickly introduce field theory models of Dark Energy so that all readers can readily relate to the discussion that follows.

Discussions of field theory models for dark energy and connections to particle physics were discussed in detail earlier by one of the current authors\cite{Singh}.
It was noted in that article that these fields must have very light mass scales in order to be cosmologically relevant today.
Discussions of realistic particle physics models with particles of light masses capable of generating interesting cosmological consequences
have been carried out by several authors\cite{GHHK}. It has been
pointed out that the most natural class of models in which to realize these ideas are models of neutrino masses with 
Pseudo Nambu Goldstone Bosons (PNGB's). The reason for this is that the mass scales associated with such
models can be related to the neutrino masses, while any tuning that needs to be
done is protected from radiative corrections by the symmetry that gave rise
to the Nambu-Goldstone modes\cite{'thooft}.

Holman and Singh\cite{HolSing} studied the finite temperature behavior of the
see-saw model of neutrino masses and found phase transitions in this model
which result in the formation of topological defects.  In fact, the critical
temperature in this model is naturally linked to the neutrino masses.

In a previous paper \cite{DarkEnergyCollapseAndBHs}. we have demonstrated the collapse of Dark Energy fields and the formation of Super Massive Black Holes. Thus it now becomes a pressing issue to investigate the other physical consequences of the collapse of Dark Energy fields. Given that the primary interactions of Dark Energy fields with the rest of the Universe are gravitational, it is particularly interesting to investigate the gravitational wave signals emitted during the process of the collapse of Dark Energy fields. This is the focus of the current work described in this paper.

In our next section, we describe the formalism and evolution equations for studying the
gravitational dynamics of dark energy field configurations. We will start by writing down the equations for
fields with any general potential. This keeps the initial discussion general. However, in light of the above discussion
on realistic models with light masses we will soon specialize to PNGB models 
which have a potential with simple and explicit form. Thus, for this purpose we will take the simplest PNGB potential\cite{GHHK}.

We then turn to a discussion of the numerical computation of the evolution of the dark energy fields coupled to gravity and the extraction of the gravitational wave signals. Finally, we describe the results of our numerical computations and the gravitational wave signals produced as a result of the collapse of the dark energy fields.

\section{  Evolution of the Dark Energy Fields in the Presence of Gravity}

We now quickly re-visit the study of the gravitational dynamics of Dark Energy field configurations.
The dynamics of fields in cosmological space-times has been extensively discussed
elsewhere (see e.g. Kolb and Turner\cite{rockymike} ). Likewise, gravitational collapse
in the context of general relativity has also been extensively discussed elsewhere 
(see e.g. Weinberg\cite{weinberggr}). These ideas can be pulled together to write 
down the evolution equations describing the coupled dynamics of the field configurations and
space-time interacting with each other. In this section we discuss the evolution equations and their solutions. In the following sections we
will discuss the extraction of gravitational waves from a collapsing star and describe the results we have obtained.

We use the 3+1 BSSN formalism to numerically study the time evolution of scalar fields in the presence of gravity.
The formalism for doing this has been previously described by Balakrishna et.al.\cite{BalakrishnaEtAl}

\subsection{The Evolution Equations} 
\label{sec:evoleqns}

The action describing a self-gravitating complex scalar field in a 
curved spacetime is:

\begin{eqnarray}
I = \int d^4 x \sqrt{-g} \left( \frac{1}{16 \pi}R \, 
    -\frac{1}{2} [ g^{\mu \nu} 
    \partial_{\mu} \Phi^* \, \partial_{\nu} \Phi  
    + V(|\Phi|^2) ]  \right) 
\label{action}
\end{eqnarray}

\noindent where $R$ is the Ricci scalar, $g_{\mu\nu}$ is the metric 
of the spacetime, $g$ is the determinant of the metric, $\Phi$ is the 
scalar field, $V$ its potential. Varying this action 
leads to equations of motion for the entire system.
Variation with respect to 
the scalar field leads to the Klein-Gordon equation for the scalar 
field

\begin{equation}
\Phi^{; \mu} {}_{;\mu} - \frac{dV}{d|\Phi|^2}\Phi = 0 . 
\label{klein-gordon}
\end{equation} 

\noindent When the variation of Eq.\ (1) is made with respect to the metric 
$g^{\mu\nu}$, we get the Einstein's equations $G_{\mu\nu}= 8\pi T_{\mu\nu}$ . The resulting stress energy tensor is:

\begin{equation}
T_{\mu \nu} = \frac{1}{2}[\partial_{\mu} \Phi^{*} \partial_{\nu}\Phi +
\partial_{\mu} \Phi \partial_{\nu}\Phi^{*}] -\frac{1}{2}g_{\mu \nu}
[\Phi^{*,\eta} \Phi_{,\eta} + V(|\Phi|^2))].
\label{stress-energy}
\end{equation}

To get numerical solutions it is convenient to use the 3+1 decomposition of Einstein's equations, 
for which the line element can be written as

\begin{equation}
   d s^{2}  = - \alpha^2  dt^2 + \gamma_{ij}  (dx^i + \beta^{i} dt) (dx^j 
              + \beta^{j} dt)  
\label{line-element}
\end{equation}

\noindent
where $\gamma_{ij}$ is the 3-dimensional metric. The latin 
indices label the three spatial coordinates. The functions $ \alpha $ 
and $ \beta^{i}$ in Eq. (\ref{line-element}) are gauge 
parameters, known as the lapse function and the shift vector respectively. 
The determinant of the 3-metric is $\gamma$ . The Greek indices run from 0 to 3 and the 
Latin indices run from 1 to 3. 

For the purpose of doing numerical evolution, the Klein-Gordon equation can be written as a first-order system.
This  is done by first splitting the scalar field into the real and imaginary parts as: 
$\Phi = \phi_1 + i\phi_2$, and then defining the following variables in terms 
of combinations of their derivatives:  $\Pi = \pi_1 + i \pi_2$ and $ 
\psi_a = \psi_{1a} + i \psi_{2a}$. Here $ \pi_1= (\sqrt{\gamma}/\alpha) 
(\partial_t \phi_1 - \beta^c \partial_c \phi_1) $ and  
$\psi_{1a}=\partial_a \phi_1$ and similarly we can replace the subscript $1$ with $2$ to get the remaining quantities of interest. With this 
notation the evolution equations become 

\begin{eqnarray}
 \partial_t \phi_1 &=&  \frac{\alpha}{\gamma^{\frac{1}{2}}} \pi_1 + 
\beta^j \psi_{1j} \\\nonumber
 \partial_t \psi_{1a} &=& \partial_a( \frac{\alpha}{\gamma^{\frac{1}{2}}} 
\pi_1 + \beta^j \psi_{1j}) \\\nonumber
 \partial_t \pi_1 &=& \partial_j (\alpha \sqrt{\gamma} \phi_1^j) 
  - \frac{1}{2} \alpha  \sqrt{\gamma} \frac{\partial V}{\partial 
  \vert \Phi \vert^2} \phi_1 \label{FirstOrderKleinGordon}
\end{eqnarray} 

\noindent
and again, we can replace the subscript $1$ with $2$ to get the remaining quantities of interest.
On the other hand, the geometry of the spacetime is evolved 
using the BSSN formulation of the 3+1 decomposition. According to this 
formulation, the variables to be evolved are 
$\Psi = \ln(\gamma_{ij} \gamma^{ij})/12$, 
$\tilde{\gamma}_{ij} = e^{-4\Psi}\gamma_{ij}$, $K = \gamma^{ij}K_{ij}$, 
$\tilde{A}_{ij}=e^{-4\Psi}(K_{ij}-\gamma_{ij} K/3)$ and 
the contracted Christoffel symbols 
$\tilde{\Gamma}^{i}=\tilde{\gamma}^{jk}\Gamma^{i}_{jk}$,
instead of the ADM variables $\gamma_{ij}$ and $K_{ij}$. The 
equations for the BSSN variables are described in Refs. \cite{bssn, 
stu}:

\begin{eqnarray}
\partial_t \Psi &=& - \frac{1}{6} \alpha K \label{BSSN-MoL/eq:evolphi}\\
\partial_t \tilde{\gamma}_{ij} &=& - 2 \alpha \tilde{A}_{ij}
\label{BSSN-MoL/eq:evolg} \\
\partial_t K &=& - \gamma^{ij} D_i D_j \alpha  + \alpha \left[
        \tilde{A}_{ij} \tilde{A}^{ij} + \frac{1}{3} K^2 + \frac{1}{2}
        \left( -T^{t}{}_{t} + T \right) \right]
\label{BSSN-MoL/eq:evolK}\\
\partial_t \tilde{A}_{ij} &=& e^{-4 \Psi} \left[
 - D_i D_j \alpha + \alpha \left( R_{ij} - T_{ij} \right) \right]^{TF}
\noindent\\
&& + \alpha \left( K \tilde{A}_{ij} - 2 \tilde{A}_{il}\tilde{A}_j^l\right) \label{BSSN-MoL/eq:evolA}\\\frac{\partial}{\partial t}\tilde \Gamma^i
&=& - 2 \tilde A^{ij} \alpha_{,j} + 2 \alpha \Big(\tilde \Gamma^i_{jk} \tilde A^{kj}                              \nonumber \\
&& - \frac{2}{3} \tilde \gamma^{ij} K_{,j}- \tilde \gamma^{ij} T_{j t} + 6 \tilde A^{ij} \phi_{,j} \Big)
                                                                \nonumber \\
&& - \frac{\partial}{\partial x^j} \Big(\beta^l \tilde \gamma^{ij}_{~~,l}- 2 \tilde \gamma^{m(j} \beta^{i)}_{~,m}+ \frac{2}{3} \tilde \gamma^{ij} \beta^l_{~,l} \Big) .
\label{BSSN-MoL/eq:evolGamma2}
\end{eqnarray}

\noindent where $D_i$ is the covariant derivative in the spatial
hypersurface, $T$ is the trace of the stress-energy 
tensor~(\ref{stress-energy}) and the label $TF$ denotes the trace-free part 
of the quantity in brackets.

The above equations are true for any general potential $V$. 
One can of course write down the corresponding equations for PNGB fields.
All we need to do is  specify the appropriate potential.
In our case, the field is a real scalar field.
The simplest potential one can write down for the physically motivated PNGB fields \cite{GHHK} can be written in the form:

\begin{equation}
V = m^4 \left[ K - \cos(\frac{\Phi}{f} ) \right]
\end{equation}

As discussed in \cite{GHHK} m is of order the neutrino mass and K is of order 1. For the sake of definiteness, in what follows we will choose $K = 1 $. We will consider such a potential for studying the dynamics in the next section.

\section{Gravitational Wave Extraction }

The evolution equations described above can be solved numerically to study the gravitational collapse of the field configurations and for the extraction of gravitational waves. 

To obtain the numerical solutions discussed below we have used the publicly available Einstein Toolkit\cite{EinsteinToolkit}.
The code uses the Method of Lines to do the time evolution. In particular, we have used the Iterative Crank Nicholson method to do the time evolution.

Guided by the evolution equations given in the previous section we define dimensionless quantities such that the field is measured in units of $f$ and time and space are measured in units of $\frac{f}{m^2}$. The energy density is measured in units of $m^4$.

 The initial conditions for the dark energy field are given below
  
  \begin{equation}
 \Phi(r,\theta, \phi)= \phi(r) [1+ \epsilon Re ( Y_{20} (\theta,\phi) )]
 \end{equation}
 \begin{equation}
  \phi(r)= \pi [1-\tanh(r-r_{0})]	  
 \end{equation}
 \begin{equation}
 	Re ( Y_{20} (\theta,\phi) ) = \frac{1}{4} \sqrt{\frac{5}{\pi}} (3\cos^2(\theta)-1)
 \end{equation}

where it should be noted that $r_0$ is a free parameter reflecting a range of possible initial conditions and the field $\Phi$ is now being measured in terms of its natural unit $f$.

We are interested in the gravitational wave signal produced by the collapse of the dark energy field configurations.
This can be extracted numerically and we have used the publicly available Einstein Toolkit for this purpose.
%\cite(ET).

 Here we extract the gauge-invariant, odd and even perturbations. 
Background  material on this can be found in Refs. \cite{Camarda99, Rezzolla99, Baker2000}.

The underlying assumption is that “far away” from the source in the wave-zone, the spacetime, or more specifically the gravitational wave-signal, can be described in terms of linear perturbations around a Schwarzschild  background  metric.   Upon  knowledge  of  the  perturbation coefficients within the numerical simulation, one can readily obtain a waveform via gauge-invariant
odd-parity (or axial) current multipoles ${Q}^{\times}_{\ell m}$ and even-parity (or polar) mass multipoles ${Q}^{+}_{\ell m}$ of the metric perturbation .The problem is then to determine the perturbation coefficients relating the  numerically  obtained  spacetime  in  the  wave-zone  to  a  perturbed  Schwarzschild  background
A particular methodology was originally developed by Regge, Wheeler
\cite{Regge1957} and Zerilli \cite{Zerilli1970b}.
respectively for examining the gravitational radiation in
terms of odd and even multipoles in the far-field of the
source. Later on, Moncrief provided a gauge-invariant
approach \cite{Moncrief1974}. (see \cite{nagar:05}.for a review).
At large distances from the source, the Gravitational waves can be
considered as a linear perturbation to a fixed background and we can write
\begin{equation}
g_{\mu\nu}=g^0_{\mu\nu}+h_{\mu\nu}\,,
\end{equation}
where $g^0_{\mu\nu}$ is the fixed background metric
and $h_{\mu\nu}$ its linear perturbation.
The background metric $g^0_{\mu\nu}$ is
usually assumed to be of Minkowski or Schwarzschild form, which we can write as
\begin{equation}
ds^2=-N dt^2 + A dr^2 + r^2(d\theta^2 + \sin^2\theta d\phi^2)\,.
\end{equation}

We can split the spacetime into timelike ,radial
and angular parts which in turn will help us in decomposing
the metric perturbation $h_{\mu\nu}$ into odd
and even multipoles, i.e.,~we can write
\begin{equation}
h_{\mu\nu}=\sum_{\ell m}\left[\left(h_{\mu\nu}^{\ell m}\right)^{(o)} + \left(h_{\mu\nu}^{\ell m}\right)^{(e)}\right]\ .
\end{equation}
The behavior of odd and even multipoles under parity transformation defines them
 $(\theta,\phi)\rightarrow(\pi-\theta,\pi+\phi)$.
Odd multipoles transform as $(-1)^{\ell+1}$ while even multipoles transform as $(-1)^\ell$.
It is always possible to expand these components in their vector and tensor spherical harmonics (e.g., \cite{thorne:80}).

By using the Hamiltonian of the perturbed Einstein equations in its ADM form \cite{Arnowitt62}, 
we can derive variational principles for the odd and 
even-parity perturbations \cite{Moncrief1974}. to 
give equations of motions that look similar to wave equations with a scattering potential.

The solutions these wave equations formed by odd-even parity perturbations are given by
the Regge-Wheeler-Moncrief and the Zerilli-Moncrief master functions, respectively.
The odd-parity Regge-Wheeler-Moncrief function reads
\begin{eqnarray} \label{eq:Qodd}
Q^{\times}_{\ell m} &\equiv& \sqrt{\frac{2(\ell+1)!}{(\ell-2)!}}
	\frac{1}{r}\left(1-\frac{2M}{r}\right) \nonumber \\
	& & \left[(h_{1}^{\ell m})^{({\rm o})}+\frac{r^2}{2} \partial_r
	\left(\frac{(h_2^{\ell m})^{({\rm o})}}{r^2}\right)\right]\ ,
\end{eqnarray}
and the even-parity Zerilli-Moncrief function reads
\begin{eqnarray} \label{eq:Qeven}
{Q}^{+}_{\ell m} \equiv \sqrt{\frac{2(\ell+1)!}{(\ell-2)!}}
\frac{r q_1^{\ell m}}{\Lambda\left[r\left(\Lambda-2\right)+6M\right]} \,,
\end{eqnarray}
where $\Lambda=\ell(\ell+1)$, and where
\begin{equation}
\label{q1}
q_1^{\ell m}  \equiv r\Lambda\kappa_1^{\ell m} + \frac{4r}{A^2}\kappa_2^{\ell m} \,,
\end{equation}
with
\begin{eqnarray}
\label{kappa1}
\kappa_1^{\ell m} & \equiv & K^{\ell m}+\frac{1}{A}\left(r\partial_r G^{\ell m}-
	\frac{2}{r}(h_1^{\ell m})^{({\rm e})}\right)\ ,\\
\label{kappa2}
\kappa_2^{\ell m} & \equiv &\frac{1}{2}\left[A H_2^{\ell m}-
	\sqrt{A} \partial_r \left(r \sqrt{A} K^{\ell m}\right)\right]\, .
\label{def:q1}
\end{eqnarray}

These master functions depend entirely on the spherical part of the metric given by the coefficients $N$ and
$A$, and
the perturbation coefficients for the
individual metric perturbation components $(h_{1}^{\ell m})^{({\rm o})}$, $(h_{2}^{\ell m})^{({\rm o})}$,
$(h_{1}^{\ell m})^{({\rm e})}$, $(h_{2}^{\ell m})^{({\rm e})}$, 
$H_0^{\ell m}$, $H_1^{\ell m}$, $H_2^{\ell m}$, $K^{\ell m}$, and $G^{\ell m}$ which
can be obtained
from any numerical spacetime by projecting out the Schwarzschild or Minkowski
background \cite{Camarda:1998wf}.
For example, the coefficient $H_2^{\ell m}$ can be obtained via
\begin{equation}
H_2^{\ell m}=\frac{1}{A}\int (g_{rr}-A) Y_{\ell m}\,d\Omega\,,
\end{equation}
where $g_{rr}$ is the radial component of the numerical metric represented in the spherical-polar coordinate basis,
$Y_{\ell m}$ are spherical harmonics, and $d\Omega$ is the surface line element of the $S^2$ extraction sphere.
The coefficient $A$ represents the spherical part of the background metric
and can be obtained by projection of the numerical metric component $g_{rr}$ on $Y_{00}$ over the extraction sphere
\begin{equation} \label{eq:H2lm}
A=\frac{1}{4\pi}\int g_{rr} d\Omega\,.
\end{equation}
Similar expressions hold for the remaining perturbation coefficients.

The odd- and even-parity master functions Eq.~(\ref{eq:Qodd}) and Eq.~(\ref{eq:Qeven}) can be
straight-forwardly related to the gravitational-wave strain and are given by
\begin{eqnarray}
\label{eq:h-Q}
h_{+}-\mathrm{i}h_{\times}&=&\frac{1}{\sqrt{2}r}\sum_{\ell,m}\left(
	Q^{+}_{{\ell m}} - \mathrm{i}\int_{-\infty}^{t}
	Q^{\times}_{{\ell m}}(t')dt'\right) \nonumber \\
	& & \;_{_{-2}}Y^{{\ell m}}(\theta,\phi)
	+ {\cal O}\left(\frac{1}{r^2}\right)\, ,
\end{eqnarray}
where ${_{-2}}Y^{{\ell m}}(\theta,\phi)$ are the spin-weight $s=-2$
spherical harmonics.

\section{Results and Conclusions}

We now present our results from the 3D numerical study of dark energy fields collapsing to produce gravitational waves.
We have used the 3+1 BSSN formalism for doing the numerical evolution.
The Evolution equations we described earlier were solved numerically using the publicly available Einstein Toolkit\cite{EinsteinToolkit}.
In particular, we extracted the gravitational wave signals emitted as a result of the collapse of dark energy field configurations.
Thus, we get the output as the odd and even Q{\scriptsize  lm} of the gravitational waves. We display these results in terms of graphs as shown below. We note, in particular, that the time period of the gravitational waves produced is comparable to the timescale of the gravitational collapse. 
 \\

 The plots of the Q{\scriptsize  lm} with time for different parameters is given in FIG. 1 to FIG. 3.
It should be noted that the units of time are given by $\frac{f}{m^2}$.
It should also be noted that $\frac{f}{m^2}$  is the fundamental timescale of the dynamics as determined by the evolution equations.

To convert into physical units, we note the following. 

The scale $f$ is the high energy symmetry breaking scale in PNGB models. In the see-saw model of neutrino masses\cite{Singh} this corresponds to the heavy scale of symmetry breaking. While $f$ has a range of possible values, the typical value of $f$ in the see-saw model of neutrino masses is $f \sim 10^{13} GeV$. The typical value of $m$ is given by $m \sim 10^{-3} eV $. It should also be noted that so far we have been working in the Particle Physics and Cosmology units in which $ \hbar=c=k=1 $ . It is straightforward to convert from these units into more familiar units using standard conversion factors\cite{rockymike}. Thus, $ 1 GeV^{-1} = 1.98 \times 10^{-14} cm$ and $ 1 GeV^{-1} = 6.58 \times 10^{-25} sec$. 

Using these conversion factors we see that the fundamental time scale of the dynamics corresponding to $\frac{f}{m^2}$ is $2 \times 10^5 years$

 We note, in particular, that the time period of the gravitational waves produced is comparable to the fundamental time scale of the dynamics. 

%\newpage

\noindent\rule{13.7cm}{0.4pt}
\newline

\centerline{\bf ACKNOWLEDGEMENTS}

This work was supported in part by a research grant from L.N.Mittal I.I.T. We thank Himanshu Kharkwal for helpful discussions and collaboration on the early part of this work.

\frenchspacing

\noindent\rule{13.7cm}{0.4pt}

% \newpage

 \newpage

% figures follow

\begin{figure*}
	\centering
	\includegraphics[width=\textwidth]{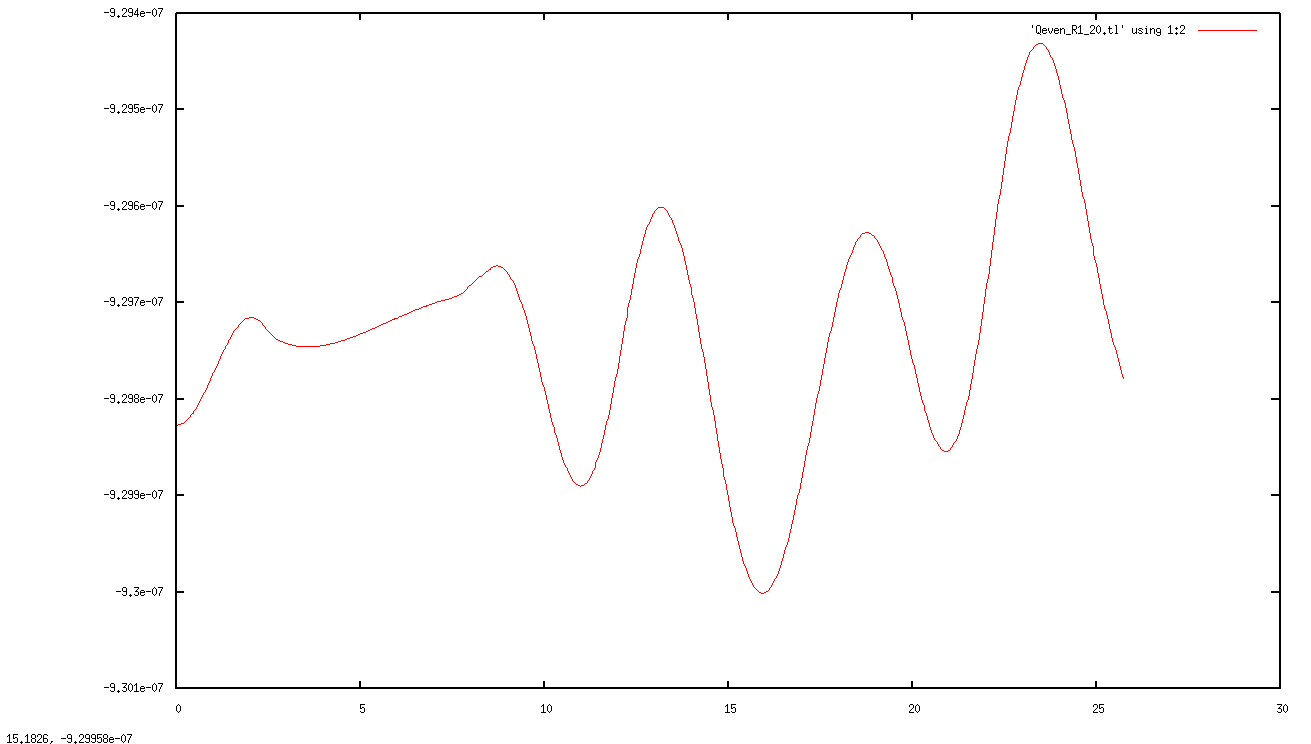}
	\caption{Even Q{\scriptsize lm} with $  l=2, m=0, \epsilon=0.0001$}
	\label{fig:ep00001w00r04fp1e4y20qeven}
\end{figure*}

\begin{figure*}
	\centering
	\includegraphics[width=\textwidth]{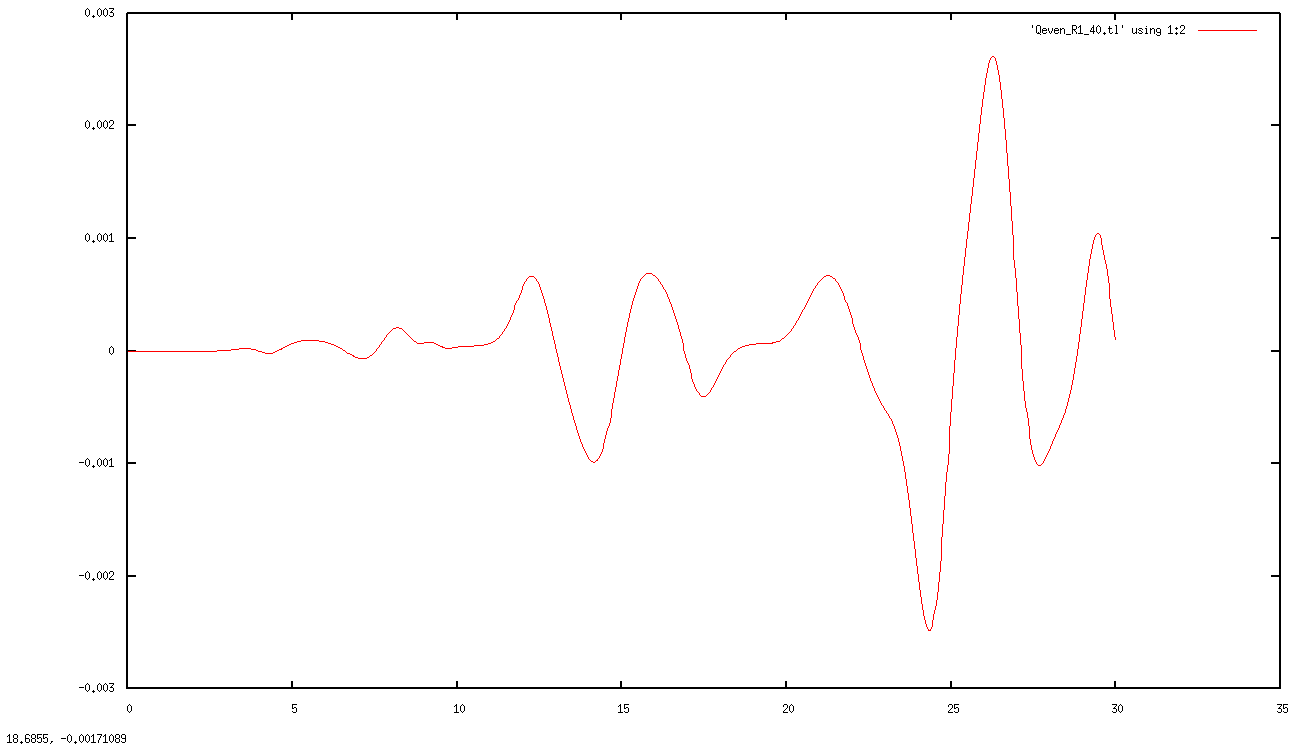}
	\caption{ Even Q{\scriptsize lm} with $  l=4, m=0, \epsilon=0.0001 $}
	\label{fig:ep00001w00fp1e5r6y20qeven40}
\end{figure*}

\begin{figure*}
	\centering
	\includegraphics[width=\textwidth]{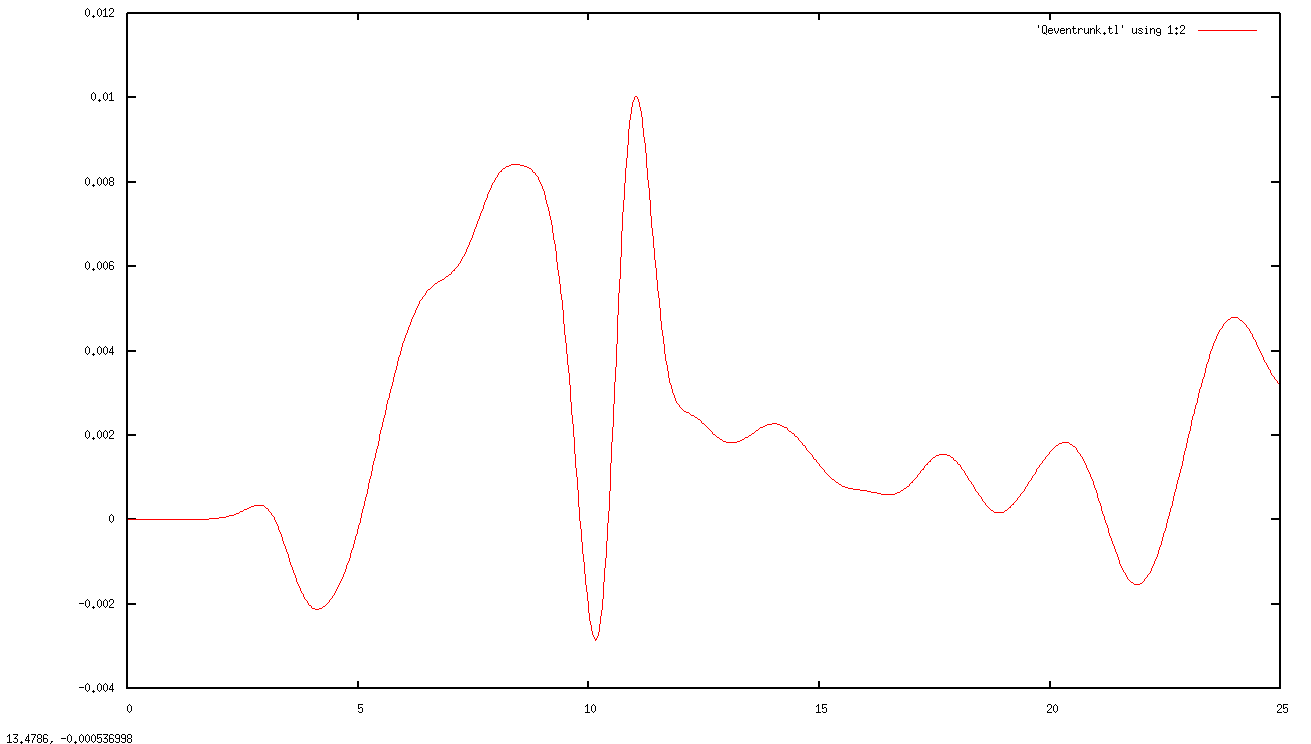}
	\caption{Even Q{\scriptsize lm} with $  l=2, m=0, \epsilon=0.01 $}
	\label{fig:ep001w00fp1e4r6y20trunkqeven}
\end{figure*}

\end{document}